\documentclass[10pt]{article}

\usepackage{amsmath,amssymb}

\usepackage[margin=2cm]{geometry}
\usepackage[utf8x]{inputenc}

\usepackage{textcomp,marvosym}

\usepackage{cite}

\usepackage{nameref,hyperref}

\usepackage[table]{xcolor}

\usepackage{array}

\usepackage{lastpage,graphicx}
\usepackage{epstopdf}

\usepackage{xcolor}

\usepackage{graphicx}

\def\s#1{\mbox{\boldmath $#1$}}

\begin{document}

\begin{flushleft}
{\Large
\textbf\newline{A visualization tool to explore alphabet orderings for the Burrows-Wheeler Transform}
}
\newline
\\
Lily Major\textsuperscript{1*},
Dave Davies\textsuperscript{1},
Amanda Clare\textsuperscript{1$\dagger$},
Jacqueline W. Daykin\textsuperscript{1,2,3$\dagger$},
Benjamin Mora\textsuperscript{4$\dagger$},
Christine Zarges\textsuperscript{1$\dagger$}
\\
\bigskip
\textbf{1} Department of Computer Science, Aberystwyth University, Aberystwyth, Ceredigion, UK
\\
\textbf{2} Department of Information Science, Stellenbosch University, South Africa
\\
\textbf{3} Univ Rouen Normandie, INSA Rouen Normandie, Université Le Havre Normandie, Normandie Univ, LITIS UR 4108, F-76000 Rouen, France
\\
\textbf{4} Computer Science Department, Swansea University, Bay Campus, Swansea, UK
\\
\bigskip
* jam86@aber.ac.uk\\
$\dagger$ These authors contributed equally to this work.

\end{flushleft}

\section*{Abstract}
The Burrows-Wheeler Transform (BWT) is an efficient invertible text transformation algorithm with the properties of tending to group identical characters together in a run, and enabling search of the text.
This transformation has extensive uses 
particularly in lossless compression algorithms, indexing, and within bioinformatics for sequence alignment tasks.
There has been recent interest in minimizing the number of identical character runs ($r$) for a transform and in finding useful alphabet orderings for the sorting step of the matrix associated with the BWT construction.
This motivates the inspection of many transforms while developing algorithms.
However, the full Burrows-Wheeler matrix is $O(n^2)$ space and therefore very difficult to display and inspect for large input sizes.
In this paper we present a graphical user interface (GUI) for working with BWTs, which includes features for searching for matrix row prefixes, skipping over sections in the right-most column (the transform), and displaying BWTs while exploring alphabet orderings with the goal of minimizing the number of runs.
% 156 words

\section*{Introduction}

The Burrows-Wheeler Transform (BWT) is an invertible permutation technique that was introduced in the context of text compression \cite{ARTICLE:BWT}. 
It is a text transformation algorithm that tends to cluster identical characters together in the output.
The transform is formed by rotating the input text until all rotations have been formed, then sorting the rotations lexicographically. The output of the BWT is the last column of the resulting matrix of sorted rotations, together with an index to the input in the matrix.
The BWT is an important transformation as it can be inverted efficiently \cite{ARTICLE:BWT} and it may be searched for any pattern $\s{p}$ in time proportional to the length of $\s{p}$ \cite{10.1145/1082036.1082039}.
It is the basis for compressors such as Bzip2 \cite{bzip2}, and is used within bioinformatics for sequence indexing and alignment in programs such as Bowtie2 \cite{ARTICLE:bowtie2}, BWA \cite{10.1093/bioinformatics/btp324}, and SOAP2 \cite{10.1093/bioinformatics/btp336}.
Additional applications of the BWT are extensive \cite{Adjeroh2008}, including shape analysis in computer vision, machine translation, and joint-source channel coding.

Graphical user interfaces to aid researchers who study the BWT are limited in functionality.
There already exist numerous websites that can display a BWT. However, these sites do not always produce the correct result.
Typically, implementations of the BWT use a special symbol to denote the end of a cyclic rotation of the input, known as the end marker. Often this is denoted as \$ and usually this is chosen to be a character that is least in the alphabet, or is less than any other character in the string to be transformed.
Assumptions that the default end marker does not appear in the input do not always hold and changing the end marker may not be supported by the current BWT GUI tools found.
Additionally, using different alphabet orderings is not supported by the current BWT GUI tools found online.

For example, in the case of Calcoolator \cite{calcoolator}, the default end marker character of $|$ (whose ASCII value is 124) is greater than all characters A-Z, a-z, and most punctuation characters due to its position within the ASCII ordering, but is not less than the $\}$ or $\sim$ characters.
Further, if the end marker appears in the text, this necessitates changing the end marker symbol.
However, using the convention of the \$ as an end marker gives different results to that of $|$ since \$ (whose ASCII value is 36) is less than all letters a-zA-Z in ASCII ordering. 
For example, for an input of $banana$, using Calcoolator with an end marker of $|$ results in $bnn|aaa$. However, using $\$$ as the end marker results in $annb\$aa$.
For inputs which contain spaces such as $banana\ banana$, the output is incorrect when using an end marker of $\$$. Calcoolator outputs $aannnnbb\$\ aaaa$ rather than $aannnnbb\ \$aaaa$ since space (ASCII value 32) is less than \$.

There are also other online BWT calculators similar to that of Calcoolator but these provide no further functionality. Both Cachesleuth \cite{CacheSleuth} and Geocachingtoolbox \cite{GeocachingToolbox.com} provide a transform in the same way as Calcoolator with an end of file marker. However, Cachesleuth defaults to \$ as the end marker instead of $|$ as with Calcoolator and Geocachingtoolbox.
Other online examples are dCode \cite{dcode} -- which returns the BWT without an end marker, and a `key' which indicates the row in the transform where the original text occurs -- and a utility \cite{haubold} which gives the whole BWT matrix as output, using $O(n^2)$ space. This implementation also only uses the ordering of the characters in ASCII for sorting, thus space characters are sorted incorrectly before the end marker $\$$.
Another existing GUI \cite{AndreaRubbi} goes further in functionality, providing a user with the BWM for an input text and allowing a search for a query string. 
However, the BWM is given as the whole matrix and thus uses $O(n^2)$ space, which is prohibitive for large input texts. There is also no ability to change the end marker used for the BWT.

Each of these existing tools does not deal with the case where the end marker is also present in the input text. Additionally, the non-existence of a GUI for display of the BWT with sufficient speed and low memory requirements motivates the work in this paper to fill these gaps for researchers.
Of particular interest is the ability to use different alphabet orderings on the same input text to observe the effects that the order has on the BWT matrix and indirectly on the number of runs in the transformed string.
Listing all rotations of the input text in the construction of a BWT matrix is however infeasible for large inputs as this requires $O(n^2)$ space.

We use the SAIS suffix array implementation by Yuta Mori\cite{MoriSais} to compute the suffix array \cite{5582081, KO2005143}, an improved space-efficient algorithm for listing all rotations of the input string. This enables multiple BWTs with different alphabet orderings to be visualized simultaneously, thus aiding intuition in choosing heuristics for alphabet orderings \cite{major2024heuristics} for reducing the number of runs in the BWT.

Here we present a new GUI tool which can calculate and display several BWT matrices with different alphabet orderings in a space-efficient manner. 
We then demonstrate an example of how this GUI enables researchers to more directly experiment with the BWT. 
We use the GUI to explore the applicability and effect of different heuristics when aiming to reorder the alphabet in order to reduce the number of runs in the BWT.

\section*{Key definitions}

An ordered alphabet is a non-empty set $\Sigma$ of unique characters of size $\sigma$, $\Sigma$ = $\{x_1, x_2, \ldots, x_\sigma\}$, where $x_1 < x_2 < \dots < x_\sigma$. 
A string $\s{s} = s_0 s_1 \dots s_{n-1}$ over an alphabet $\Sigma$ is a finite sequence of characters with length $n = \vert\s{s}\vert$. A string is also known as text, and we denote strings in boldface and characters in italic plainface.
For instance, the string $babbcd$ is a string of length 6 using the ordered alphabet $\{a,b,c,d\}$ where $a < b < c < d$. Throughout the remainder of the paper we index from 0.

For strings $\s{s}$ and $\s{s'}$, $\s{s'}$ is said to be a cyclic rotation of $\s{s}$ if there exists two strings $\s{u}$ and $\s{v}$, where $\vert\s{u}\vert, \vert\s{v}\vert \ge 1$, such that $\s{s} = \s{u}\s{v}$ and $\s{s'} = \s{v}\s{u}$. One cyclic rotation of $babbcd$ is $abbcdb$ where $\s{u} = b$ and $\s{v} = abbcd$.

The BWT matrix (BWM) for a text is constructed by sorting all cyclic rotations of the string $\s{s}$ lexicographically based on the order of the characters within $\Sigma$. 
A special character \$ is generally selected as the `end marker' and is defined as least in $\Sigma$. 
In practice, we need to consider the individual characters of the text. For instance, if the $\$$ character is used within the text, we then select the next available byte character which is not used as the end marker.
The `end marker' is appended to $\s{s}$, making each rotation unique.
The BWT, namely the transform, is the last column of the BWM, also denoted as $L$. 
The first column of the BWM is similarly denoted as $F$.

A run is a maximal repetition of a single character, and the number of runs in the BWT is denoted as $r$. 
Minimizing $r$ enables more space-efficient indexes, for instance in bioinformatics applications \cite{zakeri2023movi, Rossi2022-da}. 
A run breaker character is a character in $L$ where $L_i = L_k \ne L_j$ and $i < j < k$; the indexes $\{i, j, k\}$ are sequential.

\section*{Methodology}

First, we discuss the functionality provided by the GUI, and then give examples of the use of the GUI. Lastly, we provide heuristics based on our exploration.
Code is written in Java 11, using JavaFX, and is available with documentation on Github (\url{https://github.com/jam86/BWT-Explorer}) as source or as a compiled jar file. The code will run on an ordinary laptop.
The GUI depends on org.json 20240205 \cite{org.json}, JavaFX Base 19.0.2.1 \cite{org.openjfx.base}, JavaFX Controls 19.0.2.1 \cite{org.openjfx.controls}, JavaFX FXML 19.0.2.1 \cite{org.openjfx.fxml}, JavaFX Graphics 19.0.2.1 \cite{org.openjfx.graphics}. A Maven file is provided with the repository.

\subsection*{Practical considerations regarding implementation of the BWT}

Our tool can automatically choose an end marker by starting with $\$$ and checking if that character is present in the input text. If the character is already used, the next available character in ASCII is tried until one can be found.
Since the end marker is always treated as least in the alphabet ordering and selected automatically, our tool does not require careful selection of a character which will be either less or greater than all characters in the input text. If no end marker can be selected from the bytes 0 - 255 then our GUI implementation cannot calculate the BWT.

By storing the BWT more efficiently, our GUI is able to calculate and display BWMs for several alphabet orderings on large texts concurrently.

\subsection*{Exploration of alphabet orderings for the BWT}

Our tool allows us to explore the ways in which alphabet orderings affect the structure of the BWM. As such, multiple BWMs may be created for the input text with varied alphabet orderings.

We provide several preset alphabet orderings:
\begin{itemize}
    \item ASCII - The extended ASCII ordering
    \item Reverse ASCII - The reverse of the ASCII ordering
    \item Least frequent - Ordering characters by their frequency in the text, least frequent first
    \item Most frequent - Ordering characters by their frequency in the text, most frequent first
    \item Chapin-Tate - A hand tuned ordering for a similar problem based around compression and the BWT\cite{inproceedings:chapinHigherCompressionFromBWT}
    \item Order of appearance - The ordering of characters as they first appear in the text
    \item Vowels first - Ordering the vowels as $a < e < i < o < u < A < E < I < O < U$. The remaining characters are ordered as in ASCII and follow the vowels.
\end{itemize}
Users may also provide their own alphabet ordering as a comma separated list of characters which appear in the text.

A window into the BWM is shown for each alphabet, enabling multiple alphabet orderings to be studied, even if the size of the input text is very large.
The size of this window may be changed at the cost of memory usage. 
The last column of the BWM (the BWT) is always shown highlighted in blue; the rows of the BWM may be shown truncated to fit if the window size is smaller than the input text length.

\subsection*{Navigation and statistics}

Navigating through the BWM is necessary for larger texts that do not fit into the BWM window. The whole BWM window may be scrolled in any direction as long as there are still columns or rows in that direction. Searching for a specific prefix in either direction or scrolling the window to a specific row are also supported.
Tracking the location of a row across multiple BWMs is achieved by highlighting a row by double clicking it, and then selecting to go to that row in all transforms. That row then becomes highlighted across all BWMs and the window scrolls to display it. A row may also be found in a single transform if previously highlighted.

We provide some statistics about each BWM: the end marker character which was used (in the case that \$ is present in the text), the original size of the text in bytes, the number of runs ($r$) of the BWT, and the length of the run-length encoded transform.

\subsection*{Input and output}

A text may be selected as an input file, this is loaded bytewise rather than per character. For example, any multi byte encoding such as UTF-8 will be split into individual bytes for the BWT rather than whole unicode codepoints. This enables loading of binary data in addition to plain text.

To facilitate easy sharing of data, the BWTs may be saved in a compressed format, preserving highlighted rows and the position of the transforms in the GUI from left to right. Caching transform data is also supported, trading disk space for speed in loading a large number orderings from a saved file.
When cached, transforms do not need to be recalculated for each alphabet ordering when loaded. Noting that the number of possible alphabet orderings is $\sigma!$, this may speed up the loading of saved data when a large number of alphabet orderings have been applied. 
However, this increases the size of the stored data by $O(n)$ -- where $n$ is the length of the input text -- for each ordering.

\section*{Results}

\subsection*{The interface}

\begin{figure}
    \centering
    \includegraphics[width=1.0\textwidth]{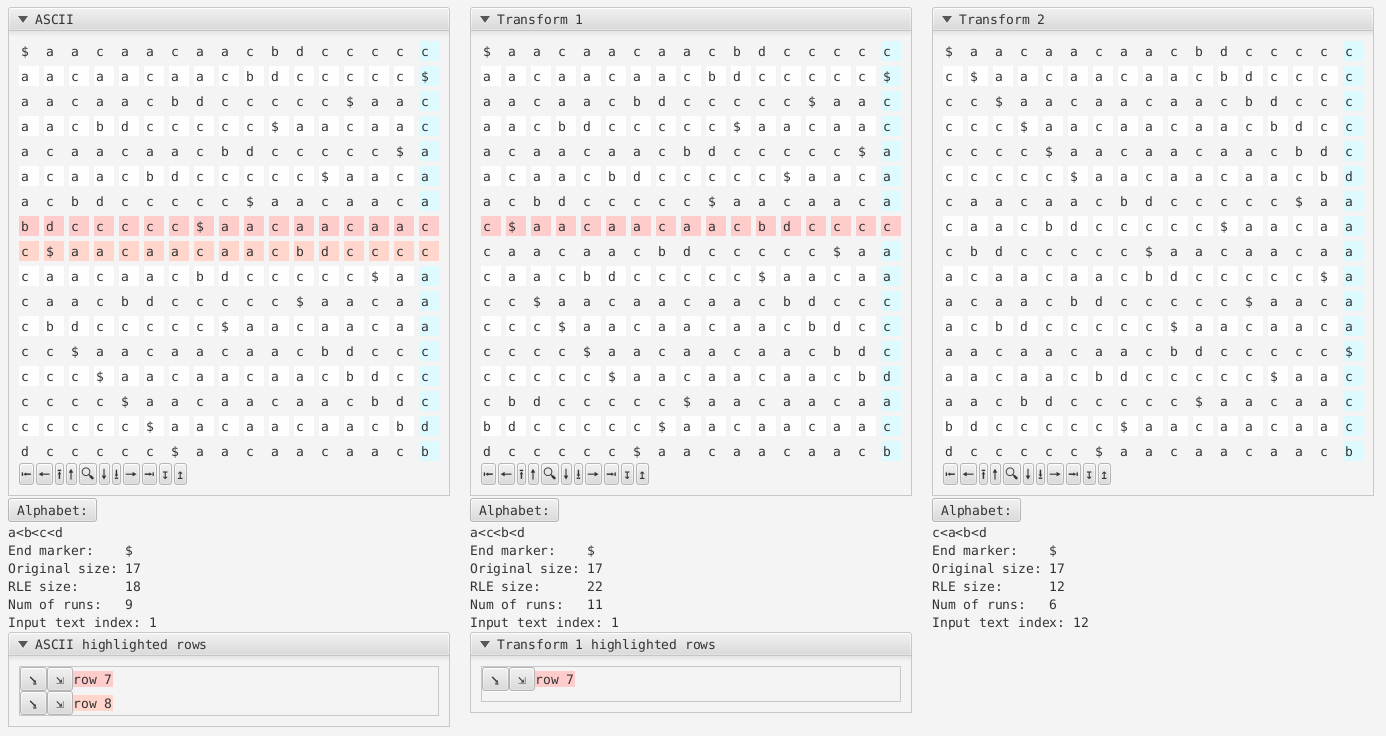}
    \caption{Three BWTs, all using $aacaacaacbdccccc$ as the input text, and each transformed using different alphabet orders. The left BWM uses ASCII ordering. In the center and right BWMs the alphabet has been re-ordered from ASCII with the goal of reducing the number of runs by joining runs of $a$ characters.}
    \label{fig:1}
\end{figure}

Figure \ref{fig:1} shows how the GUI displays three possible transforms of the same input string, each produced using a different alphabet ordering.
Navigation buttons allow scrolling the window into the BWM, although in this example the original string $aacaacaacbdccccc$ fully fits within the window size. The $L$ column is always shown fixed in blue for each BWM for any window position.
The user has chosen to highlight two rows (in pink) in the first BWM and one row in the second BWM, for further inspection.
For each of the three BWMs, information about the BWT is shown, including their alphabet ordering. For ASCII, the number of runs ($r$) in the BWT is 9. This is increased in Transform 1 to 11 and successfully decreased to 6 in Transform 2.

\subsection*{Exploring potential heuristics}

We illustrate the usefulness of the GUI by investigating through the interface the suitability of a few heuristics for reducing $r$. 
Run breaker characters increase the $r$ number and removing them is therefore desired.

The following heuristics are not exhaustive as they allow for further permutations of prioritizing blocks and other optimization choices. The goal is to rearrange matrix rows to extend runs thereby reducing $r$. 

We exemplify both heuristics using the string $\s{s} = aacaacaacbdccccc$ in Figure \ref{fig:1}.

\begin{description}

\item[H1]

Find a pair of runs in $L$, for instance visually, separated by run breaker characters. The longest potential run of characters separated by the fewest run breakers should be selected.
Out of the run breaking rows, select the row with the lowest ordering lexicographically and reorder $\Sigma$ to move this row away from the pair of runs of the same character.

When reordering, observe the characters in the BWM which determine the sort order of the rows which will become adjacent. These characters should be avoided when reordering if possible.

Repeat this procedure until all run breaking rows are moved if possible, or $r$ does not improve.

For example, in the left matrix of Figure \ref{fig:1} the ASCII ordering has a long potential run of 6 $a$ characters which are separated by two $c$ characters in $L$. We select the row beginning with $b$ (row 7) as the first to be moved and achieve this by placing $b > c$ in $\Sigma$.
This results in a worse number of runs, 11 rather than 9 (Figure \ref{fig:1}, middle matrix). 
We now select the next run-breaking row, and aim to move this.
It can be observed using our GUI that swapping $a$ and $c$ in $\Sigma$ would order our desired run of $a$s in a way that creates a longer run.
Within the rows prefixed with $a$, all $a$ characters in $L$ have $c$ as the second character. Within the rows prefixed with $c$, the $a$ characters in $L$ have either $a$ or $b$ as the second character.
The final ordering $c < a < b < d$ 
joins together the $a$ characters into a run resulting in a total of 6 runs.

This heuristic may not always improve $r$. 
The string $\s{s} = aabaaabac$ used in \cite{DBLP:journals/iandc/GiancarloMRRS23} to demonstrate alternating lexicographic order for the construction of the BWT, is not improved by this heuristic.
Fig.~\ref{fig:shortstring} shows the application of H1 to this string. 
Transform 0 (ASCII) has a run-breaker letter $b$ in row 6 breaking a potential run of 6 $a$ characters in $L$. Moving
row 6 to become row 2 in Transform 1 reorders the alphabet to $c < a < b$ and increases the run of $b$ characters but decreases the run of $a$ characters with no overall gain.

\begin{figure}[ht]
    \centering
    \includegraphics[width=0.8\textwidth]{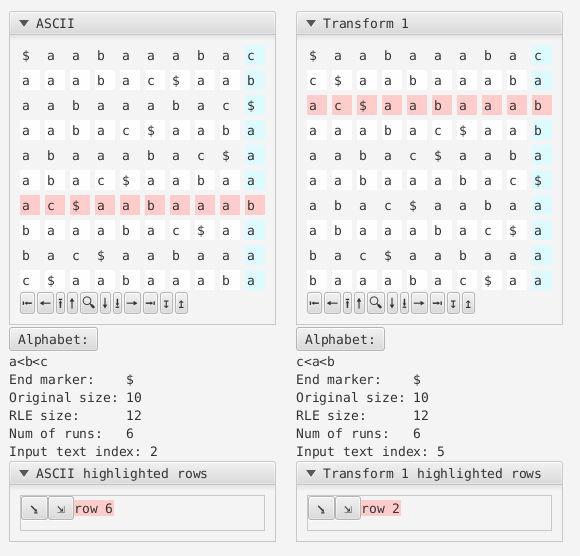}
    \caption{The string $aabaaabac$, shown with ordering $a < b < c$ in the left hand transform and $c < a < b$ in the right hand transform. For this string all permutations of the alphabet have exactly the same RLE length.}
    \label{fig:shortstring}
\end{figure}

\item[H2]

Typically, potential runs in $L$ will have some differing characters in $F$. However, the ordering of those rows with identical characters in $F$, which we call a section, can be manipulated by reordering the first characters which appear after the longest common prefix of rows. If runs of characters can be moved to either the top or the bottom of a section then longer runs may be formed by ordering two sections such that they are adjacent in the BWM.

For each section, create a list of the first characters which appear after each longest common prefix for all rows (including the prefixes of length $1$), noting the characters which appear in $L$. These are the comparisons which determine the relative ordering of rows with that common prefix. Since all rows with identical characters in $F$ can be arbitrarily ordered in the BWT, we do not consider them.
Thus, the list of character comparisons determines the local sort order for rows in a section and therefore the characters in $L$ for those rows.
 
Potential runs should be identified across sections by greedily choosing the longest potential run, breaking ties with the character least in ASCII. Partial orderings may then be created which would sort the rows with desired characters in $L$ to either the top or bottom of the sections which are to form the new run.
While sorting, using characters that also determine the sort order of other blocks should be avoided if reordering those characters would break up other runs.
It should be noted that some orderings would make joining sections impossible. If the first section sorts a desired run to the bottom and the second to the top but the ordering places the whole first section below the second then no run is formed.

From this, a list of potential partial orderings may be formed with subsets of the full alphabet ordering. An ordering may then be formed by combining these potential partial orderings if they do not have contradictory orderings, form cycles, or split up desired runs.
Characters which remain unassigned should be ordered in the new alphabet ordering as they were in the original alphabet ordering used.

In the left hand matrix in Figure \ref{fig:1} the list of compared characters in the BWM for common prefixes of length greater than 1 is $a <> b$, where $<>$ indicates that $a$ and $b$ are incomparable. This is from rows 1-3 with prefixes $aacaac$ and $aac$ and rows 4-6 with prefixes $acaac$ and $ac$. For the prefixes of length 1 the compared characters are $a <> c$ (the section beginning with $a$) and $a <> b <> c$ (the section beginning with $c$).

A run of potentially 6 $a$ characters is identified occurring between the $a$ section and the $c$ section.
In the $a$ section, the $a$ characters are already sorted to the end, relying on $c > a$. The ordering $c < a$ would place the $a$ characters at the top of the $a$ section.
In the $c$ section, the $a$ characters are already sorted to the top, relying on $(a <> b) < c$. These rows may be ordered to the bottom of the $c$ section with $(a <> b) > c$.

If $(a <> b) < c$ is chosen as an ordering, the run of $a$ characters cannot be formed due to a single row ending in $c$ at the top of the $c$ section.
Instead if $(c < a) <> (b <> d)$ is chosen, the desired run of $a$ characters is formed in $L$, reducing the number of runs from 9 to 6.

\end{description}

The heuristics H1 and H2 explored here do not improve $r$ in all scenarios, but can find better alphabet orderings in some instances. When used as part of a heuristic-driven search of the space of alphabet orderings\cite{major2024heuristics}, progress can be made towards reducing $r$. The benefits and consequences of heuristics such as these can be explored using different input texts within the GUI.

\section*{Discussion and Conclusion}

Currently available tools for visualizing and exploring the workings and outputs of BWTs are not research tools but rather exist to show how the BWT is computed in an explanatory or educational setting.
Moreover, such tools are lacking even within a teaching context; no existing tool explored by the authors includes a demonstration of backwards search for example, or offers the user any interaction with the BWM or BWT.

Our new GUI tool allows the researcher to inspect, highlight and find rows and prefixes in the BWM and to inspect and navigate runs of characters in the BWT. It also allows the researcher to alter the alphabet ordering, including intuitively, and determine the consequences.

Complex algorithms and data structures that promise to be applicable in theory may sometimes turn out not to be ideal or even not to be necessary when the actual data is inspected. For example, in bioinformatics, the popular assumption that using a de Bruijn graph data structure and finding Eulerian cycles rather than Hamiltonian cycles is the basis for modern genome assembly turns out not to be important when actual data is used, since the actual data is imperfect \cite{medvedev2021assembly}. 
The visualization of complex data, and of how algorithm and data structures behave when applied to such data is a valuable aid to researchers. 

A further example of the value of visualization in research comes from the field of computational musicology, where networks can be constructed to model melodies by associating nodes with musical notes and edges to sequences of notes. This graphical view provides a first impression of how complex a composition is and gives insights into its structure. This can guide analysis by network metrics with which to formally characterize the network and associated musical piece, thereby aiding comparison of different compositions, artists, and music genres \cite{DBLP:journals/mta/Ferretti18}.

In our example use case, any reduction in the number of runs $r$ produced in transformed strings would enable the creation of indexes with smaller footprints \cite{gagie2020fully}.
Finding an alphabet ordering that minimizes $r$ is known to be a computationally hard problem \cite{ARTICLE:complexityBWTRunsMinimizationAlphabetReordering} but making progress towards smaller $r$ with the use of heuristics is realistically achievable \cite{major2024heuristics}.
As we have demonstrated, visualization of the BWT and BWM allows us to inspect, using data examples, the types of benefit and disadvantage that different heuristics actually produce.

We have presented a new GUI for exploring BWMs and allowing different alphabet orderings.
Users of our GUI will be able to inspect different alphabet orderings for a text via a window into the BWM, vastly reducing the space required compared with enumerating all rows of the BWM.
We have also suggested heuristics that were developed from our exploration with the GUI, such as improving $r$ by moving run breaker characters, and joining potential runs by sorting rows to the top and bottom of sections in the BWM.

There are many further heuristics that could be explored in future work.
For example, context adaptive BWTs introduced in \cite{DBLP:journals/iandc/GiancarloMRRS23} induce an ordering for each common prefix but give no instructions on how to choose these orderings; the heuristics identified in our work aim to find orderings across a whole BWT but may be modified for use with the context adaptive BWT.
Additionally, the GUI could be extended to show multiple input texts and a consensus for each alphabet ordering.
Future work may also use the GUI to explore what makes a good alphabet ordering for reducing $r$ and whether this is evident from the structure of the text.

\section*{Acknowledgments}
This work is supported by the UKRI AIMLAC CDT, \url{http://cdt-aimlac.org}, grant no. EP/S023992/1.

We thank B. Chapin for personal communications regarding the Chapin-Tate ordering mentioned in the methodology.

\bibliography{main}

\end{document}